\documentclass[11pt]{article}

\usepackage{amssym}
\usepackage{times}
\usepackage{epsfig,times} 
\usepackage{wrapfig}
\makeatletter
\renewcommand{\section}{\@startsection{section}{1}{0in}
	{0.4\baselineskip}{0.1\baselineskip}{\Large\bf}}
\renewcommand{\subsection}{\@startsection{subsection}{2}{0in}
	{0.25\baselineskip}{-\baselineskip}{\large\bf}}
\renewcommand{\subsubsection}{\@startsection{subsubsection}{3}{0in}
	{0.1\baselineskip}{-\baselineskip}{\normalsize\bf}}
\makeatother
\newcommand{\fullcircle}\bullet
\newcommand{\opencircle}\circ
\newcommand{\opentriangledown}\bigtriangledown

\begin{document}

%
\thispagestyle{myheadings}
%
\markright{OG 4.3.10}
\begin{center}
%
{\LARGE \bf The energy spectra of TeV sources measured with the Durham 
Mark 6 Telescope}
\end{center}

\begin{center}
%
%
{\bf  P.M. Chadwick, K. Lyons, T.J.L. McComb, K.J. Orford, J.L. Osborne, S.M.
Rayner, S.E. Shaw, and K.E. Turver\\}
{\it Dept. of Physics, Rochester Building, Science
Laboratories, University of Durham, Durham DH1 3LE, UK}
\end{center}

\begin{center}
{\large \bf Abstract\\}
\end{center}
\vspace{-0.5ex}
%
%

A programme of detailed simulations of the response of the Durham Mark 6 
atmospheric Cherenkov telescope is in progress.
 The effective collecting area for triggering by gamma-ray showers 
 after application of selection criteria is derived as a function of energy. 
An initial result from the larger events in the 1996 and 1997
observations of the BL Lac PKS 2155--304 is that the time averaged flux
above 1.5 TeV was $(6.7\pm2.2)\times10^{-8}$ m$^{-2}$ s$^{-1}$.  

\vspace{1ex}

%
%

\section{Introduction}

The Durham Mark 6 atmospheric Cherenkov telescope, operating at
Narrabri, NSW, has so far  detected 4 sources of TeV energy
gamma-rays.  A programme of detailed simulations of the response of the
telescope to gamma-ray and cosmic ray inititiated air showers is in progress.
The aim is to improve the data analysis, refine the measurements of integral 
flux of these sources and  obtain
information on their spectra.  Some initial results of this programme are
reported here.

\section{The Simulations}
\begin{wrapfigure}[25]{l}{10.0cm}
\epsfig{file=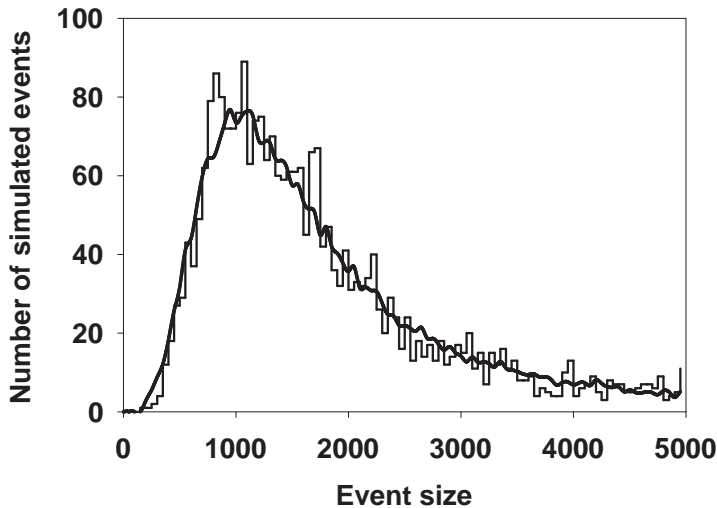,height=7.5cm,bbllx=30,bblly=310,
bburx=680,bbury=800,clip=,angle=0}
\caption{The distribution of the SIZE parameter, expressed in
digital counts, for simulated
cosmic ray showers (light line) compared with the observed distribution for
off-source cosmic rays, normalised to the same total number of events
(heavy line).
A digital counts to photoelectron ratio of 4.5 is applied to the
simulations to give this fit. \label{fig:camsum}}
\end{wrapfigure}
The Monte Carlo simulation is performed in two stages. In the first
a recent version of the MOCCA92 code (Hillas, 1995) models the shower
development in the atmosphere and generates the position, time and
direction of arrival of Cherenkov photons at the telescope mirrors (expressed
as potential photoelectrons at the PMT
photocathodes). This version includes wavelength dependent atmospheric 
absorption, mirror reflectivity and PMT quantum efficiency.  The second stage
models the mirror optics, the PMT detector arrays, the telescope 
triggering, the noise and the overall gain of the Mark 6 telescope. This produces 
images in 
terms of digital counts that can be processed in the same way as the 
real data.   

A detailed description of the Mark 6 telescope has been given by Armstrong et al.
(1999).  The properties that are incorporated into the second 
stage of the simulations are as follows.  The telescope is 260 m above 
sea level. It has three 7 m diameter f/1.0 parabolic mirrors mounted
with their centres 7 m apart on a single alt-azimuth platform.
At the focus of the central mirror there is an imaging camera
consisting of a close packed hexagonal array of 91 circular
2.5cm diameter Hamamatsu R1924 PMTs with $0.25^{\circ}$ spacing.  Conical
reflective light concentrators largely eliminate the dead area
between the tubes. The array covers a 1.3$^{\circ}$ radius field
of view.  Surrounding this is a guard ring of 18 circular 5cm diameter
Burle 8575 PMTs.  The photodetectors of the left and right
mirrors each have 19 Phillips XP3422 hexagonal PMTs covering
field of view of the 91 2.5 cm PMTs.  The mirror surface is Alanod 410G
anodised aluminium with a reflectivity $\geq$75\% in the
wavelength range 700 to 350 nm and falling to $\sim$60\% at 280 nm.
  The response of the 2.5 cm PMTs cuts
off at this wavelength.  The point spread function of the
mirrors may be represented by the sum of two Gaussians, a narrow
component with an rms radius of $0.18^{\circ}$ and a `skirt'
with an rms radius of $0.45^{\circ}$ contributing 24\% of the peak
amplitude.
The trigger requires, within a coincidence time of 10 ns, a
signal from corresponding left and right mirror PMTs and any two adjacent
centre mirror PMTs of the group of 7 that cover the same region of the sky.
A direct measure of the gain of the system by means of a radioactive
light pulser applied to each PMT gives a digital counts to photoelectron
ratio, dc/pe$\sim4$.

\begin{wrapfigure}[24]{l}{10.0cm}
\epsfig{file=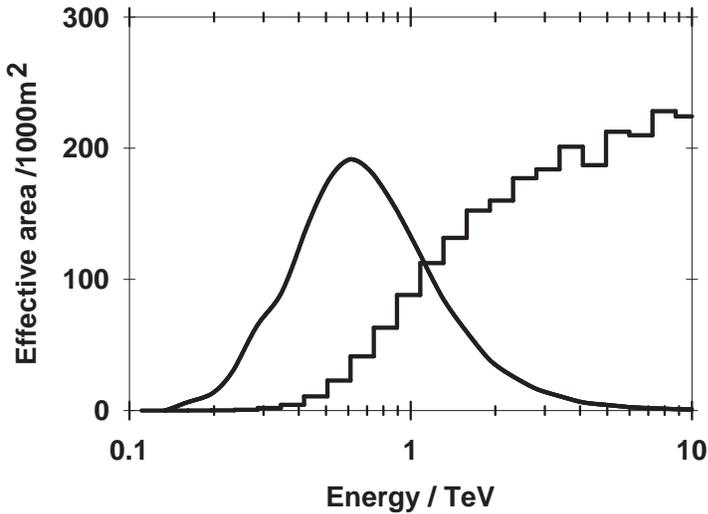,height=7.5cm,bbllx=30,bblly=310,
bburx=680,bbury=800,clip=,angle=0}
\caption{The Histogram gives the variation of the effective area with 
energy for gamma-ray shower
impact points within 300 m of the telescope. The curve, having a
vertical scale in abitrary units, shows the form of the triggering
spectrum for a power law differential source energy spectrum of index
2.6.
 \label{fig:threshplot}}
\end{wrapfigure}
Ultimately showers will be simulated over the full range of
zenith angles.  Initially a zenith angle of 30$^{\circ}$ has
been chosen as representative of the bulk of the observations.
Some dependence of the shape of the 
Cherenkov images on the
strength of the geomagnetic field component perpendicular to the
shower axis has been observed.  An azimuth angle of 180$^{\circ}$
gives 0.48 G for this component, again a typical value for many of the
observations.

In order to calibrate the gain of the telescope 31000 cosmic ray
showers with energies between 0.3 and 30 TeV were generated.
For each shower the telescope is placed at 5 random positions
within an impact radius of 300m and for each of these the shower direction
takes 4 random values out to 2$^{\circ}$ from the centre of the field
of view.  The same values of cosmic ray composition and 
spectra were adopted as were used by Mohanty et al. (1998).
By adjustment of the discriminator level in the simulations 
the typical observed off-source trigger rate at 30$^{\circ}$ zenith angle of
640 cpm was matched.  The distribution of the SIZE of 
the triggering events in units of the
number of photoelectrons in the imaging tubes was produced.
This was then compared with the observed SIZE distribution of
off-source showers  expressed in digital counts.  As can be seen    
 in figure~\ref{fig:camsum} a value of dc/pe of 4.5 gives a good
fit, in agreement with the directly measured value.

\section{Results of Gamma-Ray Simulations}

A total of 50000 gamma-ray showers with energies between 0.1 and 30
TeV were generated in order to determine the triggering probability
as a function of energy and then the probability of the gamma rays
surviving the various selection criteria which are applied to enhance 
the gamma-ray signal to cosmic ray background ratio.  Again for each
shower the telescope is placed randomly at 5 positions within an
impact radius of 300 m of the shower axis.  The shower energies were
drawn randomly from  a differential power law spectrum of index 2.4
but the exact value of this index does not matter as the triggering
probability at a given energy is given by the ratio of the number of
showers triggering to the number of showers generated at that energy.
Using the discriminator level and dc/pe ratio determined from the
cosmic ray flux the total number of triggers was 9256.  The effective
area for triggering as a function of energy is shown as the histogram in 
figure~\ref{fig:threshplot}.  This is obtained by multiplying the
triggering probabilities by the the 300 m radius target area.

Any gamma-ray source energy spectrum can be multiplied into this distribution
and integrated to obtain the corresponding total triggering rate.  The
curve in figure~\ref{fig:threshplot} shows the smoothed effective area
distribution multiplied by $E^{-2.6}$.  It can be seen that
significant triggering starts at $\sim200$ GeV.  A traditional
definition of the energy threshold is the energy of the peak of the
triggering energy spectrum, which lies at 600 GeV for the Mark 6
telescope.  The effective area
above threshold may be defined as  that of a detector with 100\%
triggering probability above the threshold energy and zero below it
that has the same total triggering rate as the telescope.  The value
of this effective area is $3.8\times10^5$ m$^2$.   

The process of gamma-ray/cosmic ray separation based on image
parameters (Hillas 1985) is then applied to the triggering events.  The first
selection is based on SIZE, the sum of the digital counts in the central camera
tubes.  Those events with SIZE below 200 dc are rejected.  `Image'
tubes are then defined, in this study, firstly as those at more than 
4.25 times the rms
sky noise and secondly those at more than 2.25 times the sky noise
that are adjacent to the first. Events with less than two image tubes
are removed. A DISTANCE parameter, the distance of
the image centroid from the centre of the camera is evaluated and
those with DISTANCE $>$1.1$^{\circ}$ or have peak brightness in the
guard ring are rejected.  These criteria result in the
removal of 40 to 50\% of the triggered cosmic ray events.  The primary
aim is to retain only those events to which useful image parameters
can be assigned but the simulations show that only $\sim20\%$ of the
gamma-rays will have been removed.  Further image parameters are then
evaluated: the WIDTH, the rms spread of the image along the minor axis;
the ECCENTRICITY, the ratio width/length; the CONCENTRATION defined as 
the fraction of dc in SIZE that is not in image tubes; and $D_{\rm
dist}$  the difference in position of the image centroids in the 
left and right detectors.  Lower values of these last two parameters
should favour the selection of gamma-ray images.  The final, and most
significant parameter is ALPHA, the angle between the image's long
axis and the line from its centroid to the source position 
(i.e. the camera centre for the
simulations).
A comparison of the ALPHA distributions for simulated gamma-rays and
observed cosmic rays is shown in figure 3.

\begin{wrapfigure}[20]{l}{10.0cm}
\epsfig{file=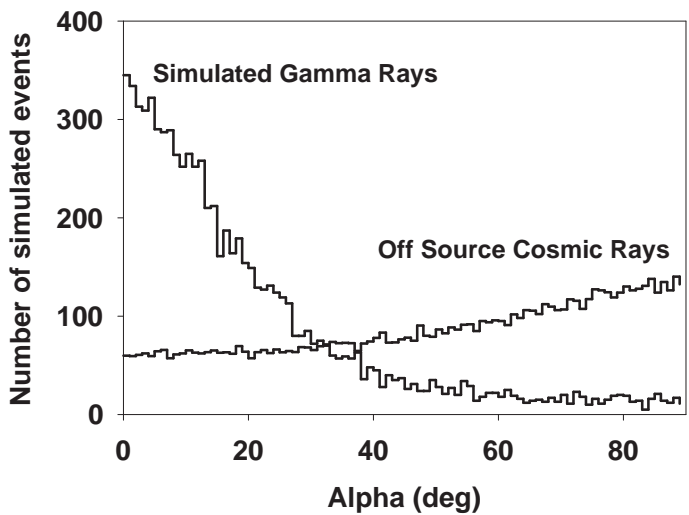,height=7.5cm,bbllx=30,bblly=310,
bburx=680,bbury=800,clip=,angle=0}
\caption{Distributions of image parameter ALPHA for simulated gamma-ray
showers and observed off-source cosmic ray showers.} 
\label{fig:alphaplot}
\end{wrapfigure}

From a consideration of earlier simulations and some empirical
optimisation, parameter cuts as listed in Table 1 plus
the requirement that ALPHA $<$ 22.5$^{\circ}$ has been adopted by the Durham
group for the analysis of data from AGN source candidates.  The SIZE
independent DISTANCE and ECCENTRICITY cuts are applied in order to
ensure well defined values of ALPHA.  By applying these cuts to the
simulated gamma-ray events we should obtain distributions of effective
area against energy which can be applied to the excess counts in each
size bin for an observed source to obtain the corresponding fluxes and thus 
an indication of its energy spectrum. At the present stage of the
work, however, we find that the fluxes obtained from the smaller
size bins are not sufficiently robust against refinements in the
simulations of noise and image blur to be reliably quoted.  For the
two larger size bins the results are more robust and we apply them in
the next section.   

\section{The Source PKS 2155--304}

Positive results from the close X-ray selected BL Lac PKS 2155--304
were obtained for the 1996 and 1997 observing seasons (Chadwick et
al., 1999) including an indication of a correlation of VHE with X-ray
fluxes.  For zenith angles less than 45$^{\circ}$ a total of 544 excess
gamma-rays were observed on-source in 32.5 hours of observation.
The the top two size bins contain an excess of 358 events, with a
significance of 3.1$\sigma$. The gamma-ray simulations indicate
that the effective area against energy for these large events when
multiplied by an $E^{-2.6}$ spectrum peaks at 1.5 TeV.  Treating this
as their `threshold energy' the effective area above threshold is
$4.6\times10^4$ m$^2$ and the corresponding flux above this energy
is $(6.7\pm2.2)\times10^{-8}$ m$^{-2}$s$^{-1}$, where the error is
statistical only.
\begin{table*}[h]
\begin{center}
\begin{tabular}{@{}lccccc}
\hline
Parameter&Ranges&Ranges&Ranges&Ranges&Ranges\\
\hline
\hline
{\it SIZE} (d.c.)&$500-800$&$800-1200$&$1200-1500$&$1500-2000$&$2000-10000$\\
{\it DISTANCE}&$0.35^{\circ}-0.85^{\circ}$&$0.35^{\circ}-0.85^{\circ}$&$0.35^{\circ}-0.85^{\circ}$&$0.35^{\circ}-0.85^{\circ}$&$0.35^{\circ}-0.85^{\circ}$\\
{\it ECCENTRICITY}&$0.35-0.85$&$0.35-0.85$&$0.35-0.85$&$0.35-0.85$&$0.35-0.85$\\
{\it WIDTH}&$ < 0.10^{\circ}$&$ < 0.14^{\circ}$&$ < 0.19^{\circ}$&$ < 0.32^{\circ}$&$ < 0.32^{\circ}$\\
{\it CONCENTRATION}&$ < 0.80$&$ < 0.70$&$< 0.70$&$ < 0.35$&$< 0.25$\\
$D_{\rm dist}$&$ < 0.18^{\circ}$&$ < 0.18^{\circ}$&$ < 0.12^{\circ}$&$ < 0.12^{\circ}$&$ < 0.10^{\circ}$\\
\hline
{\it Excess on-source} & 29 & 74 & 83 & 138 & 220 \\
{\it Off-source Events} & 227 & 371 & 433 & 1546 & 2042 \\
\hline
\end{tabular}
\end{center}
\caption{The image parameter selections applied to the PKS 2155--304
data recorded at zenith angles less than $45^\circ$ during 1996 and
1997.  The bottom two rows show the numbers of
excess on-source events compared to off-source events resulting from the 
imposition of these cuts plus ALPHA $<$ 22.5$^{\circ}$.  
\label{select_table}}
\end{table*}

\section{Conclusions}
Work will continue on the simulations of the trigger and the 
effects of noise  and the mirror blur on the images near threshold.
The systematic uncertainties in the derived fluxes need evaluating.
A refinement of the selection criteria and their energy
dependence should be possible.

We are grateful to the UK Particle Physics and Astronomy Research
Council for support of the project and to A.M. Hillas for providing an
updated version of MOCCA92.

\vspace{1ex}
\begin{center}
{\Large\bf References}
\end{center}
%
Armstrong, P. et al. 1999, Experimental Astron., in press\\
Chadwick, P.M. et al. 1999, ApJ, 513, 161\\
Hillas, A.M. 1985, Proc. 19th ICRC (La Jolla) 3, 445\\
Hillas, A.M. 1995, Proc. 24th ICRC (Rome) 1, 270\\
Mohanty, G. et al. 1998, Astrop. Phys., 9, 15\\
\end{document}